# Optical melting of the transverse Josephson plasmon: a comparison between bilayer and trilayer cuprates

W. Hu[1,*], D. Nicoletti[1], A. V. Boris[2], B. Keimer[2], and A. Cavalleri[1,3,†]

*1. Max Planck Institute for the Structure and Dynamics of Matter, Center for Free Electron Laser Science, 22761 Hamburg, Germany*
*2. Max Planck Institute for Solid State Research, 70569 Stuttgart, Germany*
*3. Department of Physics, Clarendon Laboratory, University of Oxford, OX1 3PU Oxford, United Kingdom*

**Abstract:** We report on an investigation of the redistribution of interlayer coherence in the trilayer cuprate $Bi_2Sr_2Ca_2Cu_3O_{10}$. The experiment is performed under the same apical-oxygen phonon excitation discussed in the past for the bilayer cuprate $YBa_2Cu_3O_{6.5}$. In $Bi_2Sr_2Ca_2Cu_3O_{10}$, we observe a similar spectral weight loss at the transverse plasma mode resonance seen in $YBa_2Cu_3O_{6.5}$. However, this feature is not accompanied by light-enhanced interlayer coherence that was seen in $YBa_2Cu_3O_{6+x}$, for which the transverse plasma mode is observed at equilibrium even in the normal state. These new observations offer new experimental perspective in the context of the physics of light-enhanced interlayer coupling in various cuprates.

[*] wanzheng.hu@mpsd.mpg.de
[†] andrea.cavalleri@mpsd.mpg.de

Superconducting-like c-axis coherence has been found in various cuprate superconductors after optical excitation at specific phonon resonances. For single-layer compounds like $La_{1.675}Eu_{0.2}Sr_{0.125}CuO_4$, the excitation of an in-plane Cu-O stretching phonon was shown to enhance c-axis coherent coupling [1]. Importantly, a plasma mode appeared at approximately the same frequency (~65 cm$^{-1}$) as that seen in the superconducting state of the optimally doped single layer cuprate $La_{1.84}Sr_{0.16}CuO_4$. The transient c-axis coherence in $La_{1.675}Eu_{0.2}Sr_{0.125}CuO_4$ was observed up to the charge-ordering temperature [1], and was interpreted as a sign of light-induced superconductivity [2] after optical melting of stripes [3].

Similarly, in the bilayer cuprate $YBa_2Cu_3O_{6.6}$, for which hole doping is also near 1/8, excitation of the apical oxygen phonon resulted in a strengthened interlayer superconducting coupling below the equilibrium transition temperature ($T_c$ = 61 K), and in the appearance of a plasma resonance above $T_c$ [4]. This phenomenon was directly correlated to the melting of charge order in the same material [5], in analogy with the physics of $La_{1.84}Sr_{0.16}CuO_4$.

More difficult to understand are the results at lower doping levels, in $YBa_2Cu_3O_{6.5}$ and $YBa_2Cu_3O_{6.45}$, for which the light-induced plasma edge appeared at even higher temperatures, even above the equilibrium charge ordering temperature. Light induced interlayer coupling was detected up to temperatures that are in fact rather reminiscent of the equilibrium pseudogap ($T^*$). Further experimental work in $YBa_2Cu_3O_{6.5}$ has associated the appearance of a low frequency inter-bilayer plasma mode with a partial melting of the intra-bilayer plasmon at high frequency, as if 'coherence' were transferred from the $CuO_2$ bilayer to the inter-bilayer region of the unit cell [6]. This hypothesis was qualitatively corroborated by observations made with ultrafast x-ray scattering, which indicated a rearrangement of the unit cell atoms that involve transient expansion (compression) of $CuO_2$ intra-bilayer (inter-bilayer) distances [7].

Hence, the physics of such light-induced superconductivity appears to be more general than optical melting of stripes. This notion was further confirmed by the observation of a related transient superconducting-like phase in the organic material $K_3C_{60}$ [8].

All these experimental observations are now the basis for a number of theoretical interpretations of the light-stabilized superconductivity in cuprates, including transient suppression of competing order [9,10], redistribution of phase fluctuations between inter- and intra-bilayer Josephson junctions [11], parametric cooling [12,13], electron-phonon coupling [14], and phonon squeezing [15,16], for which no comprehensive picture exists.

Here, we extend our analysis to another non-charge ordered, multilayer cuprate, a useful comparison for the data in $YBa_2Cu_3O_{6.5}$. We study the response of trilayer $Bi_2Sr_2Ca_2Cu_3O_{10}$ (underdoped, with midpoint $T_c$ = 99 K and $\Delta T_c$ =10 K), under similar driving.

We report a transient high-frequency spectral weight loss at the transverse plasma mode. However, in this compound the inter-trilayer coherence was not enhanced at any temperature and the response at the transverse plasma mode disappeared immediately above the superconducting transition temperature. Different scenarios are analyzed to rationalize this observation.

In Figure 1(a) we show the structure of $YBa_2Cu_3O_{6.5}$, which exhibits *two* $CuO_2$ planes in one unit cell. Each copper atom in the plane is coordinated with four oxygen atoms in the same layer and with one apical-oxygen atom in the perpendicular direction. In $Bi_2Sr_2Ca_2Cu_3O_{10}$, the structure contains *three* closely spaced $CuO_2$ planes and no Cu-O chains. The copper atoms in the two outer $CuO_2$ planes are five-fold coordinated with oxygen atoms as in $YBa_2Cu_3O_{6.5}$, whereas the copper atoms at the inner $CuO_2$ plane only bind to four oxygens (Fig. 1(e)). Previous studies have shown that the doping level is not homogeneous in these three planes, with the outer $CuO_2$ planes being more doped [17,18] and exhibiting larger superconducting gaps than the inner $CuO_2$ plane [18].

The c-axis equilibrium reflectivity of both samples is shown in Figure 1 (b) and (f) in the spectral range below 80 cm$^{-1}$. In $YBa_2Cu_3O_{6.5}$ the superconducting transition is evidenced by the appearance of an inter-bilayer Josephson Plasma Resonance (JPR) at ~30 cm$^{-1}$, as already reported in [4, 6]. In Bi-based cuprates, the highly anisotropic crystal structure strongly suppresses the interlayer Josephson coupling strength [19], thus resulting in typical plasma edges in the GHz frequency range. This is also the case for $Bi_2Sr_2Ca_2Cu_3O_{10}$, where signatures of a JPR are only found at $<$ 10 cm$^{-1}$.

The broadband c-axis optical conductivity and energy loss functions, measured for both samples at equilibrium, are reported in Figure 1 (c), (d), (g), (h). Both bilayer $YBa_2Cu_3O_{6.5}$ and trilayer $Bi_2Sr_2Ca_2Cu_3O_{10}$ show incoherent c-axis responses with low conductivity. Infrared-active phonons appear as peaks in $\sigma_1(\omega)$, at frequencies ranging from 100 to 650 cm$^{-1}$. These modes sharpen with decreasing temperature.

In $YBa_2Cu_3O_{6.5}$, two inequivalent apical oxygen positions result in two apical oxygen phonons around 550 and 630 cm$^{-1}$ (Fig.1(d)). A broad conductivity peak appears around 400 cm$^{-1}$ with decreasing temperature (shaded area in Fig.1(d)), causing a strong reduction in the oscillation strength of a 320 cm$^{-1}$ planar-oxygen bending phonon.

In $Bi_2Sr_2Ca_2Cu_3O_{10}$, a single apical oxygen phonon is observed at 575 cm$^{-1}$ (Fig. 1(h)) [20]. In this case, it is the spectral weight of this apical oxygen mode which gets depleted below $T_c$ and transferred to a ~500 cm$^{-1}$ peak. Similar to the 400-cm$^{-1}$ mode in $YBa_2Cu_3O_{6.5}$, this 500-cm$^{-1}$ peak in $Bi_2Sr_2Ca_2Cu_3O_{10}$ is associated with a transverse Josephson plasma resonance [21], which has been also found in various multilayer cuprates [22].

Analogous modes of non-phononic character are also found in the energy loss function of both $YBa_2Cu_3O_{6.5}$ and $Bi_2Sr_2Ca_2Cu_3O_{10}$ at ~500 cm$^{-1}$ and ~550 cm$^{-1}$, respectively (Fig. 1(c) and Fig. 1(g)). These can be identified as intra-bi(tri)layer longitudinal plasma modes.

The physics of longitudinal and transverse plasmons can be understood by considering multilayer cuprates as stack of Josephson junctions, with Cooper pairs tunneling between neighboring $CuO_2$ layers.

In the bilayer case ($YBa_2Cu_3O_{6.5}$), one can think of two Josephson junctions in one unit cell: An intra-bilayer junction between closely spaced planes and an inter-bilayer junction (Fig. 1(c)). Each junction is associated with a longitudinal Josephson plasma mode, which is detected as a peak in the loss function $-Im(1/\varepsilon(\omega))$, or, equivalently, as an edge in the optical reflectivity. In addition to these two modes, out-of-phase oscillations of the Josephson plasma within and between pairs of copper oxide layers give rise to the transverse plasma mode, revealed by a peak in the real part of the optical conductivity $\sigma_1(\omega)$ [23]. For underdoped $YBa_2Cu_3O_{6.5}$, the transverse mode, as well as the intra-bilayer longitudinal mode, develop at temperatures much higher than the superconducting transition temperature $T_c$, which is suggestive of pre-existing interlayer coherence in the normal state [24]. As shown in Fig. 1(c)-(d), both plasma modes in $YBa_2Cu_3O_{6.5}$ are clearly observed even at $T = 1.2\ T_c$, indeed almost as large as those observed deep in the superconducting phase ($T = 0.2\ T_c$).

In trilayer $Bi_2Sr_2Ca_2Cu_3O_{10}$, the physics of longitudinal and transverse plasma modes is similar to that described above. Indeed, despite the larger unit cell, one can think of the trilayer block composed of three equally spaced $CuO_2$ planes as a single Josephson junction. The tunneling dynamics is that of a single mode, with the charge in the middle $CuO_2$ plane not changing significantly in an infrared field [21,25]. Hence, one finds only two longitudinal plasma modes and a single transverse plasma mode in trilayer $Bi_2Sr_2Ca_2Cu_3O_{10}$ (Fig.1 (f)-(h)), which, unlike in $YBa_2Cu_3O_{6.5}$, all appear exactly at the superconducting transition [21].

Within the Josephson superlattice model, the transverse plasma frequency is defined by $\omega_T^2 = (d_2\omega_{Jp1}^2 + d_1\omega_{Jp2}^2)/(d_1 + d_2)$, where $d_1$ and $d_2$ are the thickness of the inter- and intra-bi(tri)layer junctions, respectively, and $\omega_{Jp1}$ and $\omega_{Jp2}$ are the corresponding longitudinal Josephson plasma frequencies.

For our pump-probe experiments, we used mid-infrared pulses centered at 15 μm (667 cm$^{-1}$) and polarized along the direction perpendicular to the $CuO_2$ planes to drive the apical oxygen phonons. The excitation pulses were generated by difference-frequency mixing in an optical parametric amplifier, with a 300 fs pulse duration and 4 mJ/cm$^2$ pump fluence. Figure 2 (a) and (d) illustrate the atomic motions in the phonon driven state for $YBa_2Cu_3O_{6.5}$ and $Bi_2Sr_2Ca_2Cu_3O_{10}$, respectively. A typical spectrum of the mid-infrared excitation pulse is plotted together with the equilibrium c-axis optical response functions in the superconducting state in Fig.2 (b), (c), (e), (f).

To probe the transient optical properties, we used broadband THz pulses generated by laser-ionized gas plasma [26], and detected by electro-optical sampling of the THz field either in a gas plasma [6, 26], or in a 50-μm-thick z-cut GaSe crystal [27]. The polarization of the THz probe pulses was also set to be perpendicular to the $CuO_2$ planes.

With this technique, we could directly determine the reflected THz field for different pump-probe time delays $t$ after photo-excitation. By separately modulating mid-infrared pump and THz probe and simultaneously recording the signals using two lock-in amplifiers, we retrieved both the pump-induced THz field change $\Delta\tilde{E}(\omega,t)$ and the stationary THz field $\tilde{E}(\omega,t)$. The complex reflection coefficient of the photo-excited sample, $\tilde{r}'(\omega,t)$, was calculated using the relation

$$\frac{\Delta\tilde{E}(\omega,t)}{\tilde{E}(\omega,t)} = \frac{\tilde{r}'(\omega,t) - \tilde{r}^0(\omega)}{\tilde{r}^0(\omega)}$$

where $\tilde{r}^0(\omega)$ is the equilibrium complex reflection coefficient obtained from steady-state optical spectroscopy. In order to take into account the pump-probe penetration depth mismatch, we used a multi-layer model [28] for $\tilde{r}'(\omega,t)$, involving a fully photo-excited layer, with thickness equal to the mid-infrared pump penetration depth (4 μm for $YBa_2Cu_3O_{6.5}$ and 6 μm for $Bi_2Sr_2Ca_2Cu_3O_{10}$) and complex refractive index $\tilde{N}^{photo-excited}(\omega,t)$, and a bottom layer in which the refractive index retains the equilibrium value $\tilde{N}^0(\omega)$. By extracting $\tilde{N}^{photo-excited}(\omega,t)$ from $\tilde{r}'(\omega,t)$ using this multi-layer model, we could retrieve the transient optical response functions of the photo-excited layer.

The pump-induced changes to the real part of the optical conductivity, $\sigma_1(\omega,t)$, measured in the superconducting state ($T < T_c$), are plotted for both samples in Fig. 3(a)-(b), for frequencies $\omega \gtrsim$ 150-250 cm$^{-1}$. For $YBa_2Cu_3O_{6.5}$, the 400-cm$^{-1}$ transverse plasma mode exhibits a prompt shift to lower frequencies upon excitation, while in $Bi_2Sr_2Ca_2Cu_3O_{10}$, the 500-cm$^{-1}$ peak is completely depleted and its spectral weight gets transferred to the apical oxygen phonon at 575 cm$^{-1}$.

These data suggest that for both cuprates pumping the apical oxygen phonon results in a weakening of the intra-bi(tri)layer Josephson coupling. The color plots in Fig. 3 (a) and (b) display the light-induced changes in $\sigma_1(\omega)$ as a function of pump-probe time delay for both samples.

As discussed previously, in $YBa_2Cu_3O_{6.5}$, the plasma mode red-shift is observed up to temperatures far above equilibrium $T_c$ (Fig. 3(c)), while for $Bi_2Sr_2Ca_2Cu_3O_{10}$, this transient spectral weight redistribution occurs only when the experiment is performed below $T_c$ (Fig. 3(d)). The relaxation dynamics for the two samples are also different: While the transient conductivity changes in $YBa_2Cu_3O_{6.5}$ relax back to equilibrium within about 7 ps, the time scale found in $Bi_2Sr_2Ca_2Cu_3O_{10}$ is faster (~3 ps).

The dynamics of the inter-(bi)trilayer coupling strength was probed in both sets of experiments by measuring the transient optical response at lower frequencies, under the same excitation conditions. Single-cycle THz probe pulses were generated using either a ZnTe crystal (for $YBa_2Cu_3O_{6.5}$) or a photoconductive antenna (for $Bi_2Sr_2Ca_2Cu_3O_{10}$). These pulses were shone onto the samples with polarization perpendicular to the $CuO_2$ planes and were then detected by electro-optical sampling in a ZnTe crystal, thus being able to probe the dynamical evolution of the optical response for $\omega \lesssim 80$ cm$^{-1}$.

As displayed in Figure 4, in the superconducting state of $YBa_2Cu_3O_{6.5}$, at the peak of the pump-probe response ($t$ = 0.8 ps), the imaginary part of the optical conductivity, $\sigma_2(\omega)$, changes slope and diverges more strongly toward low frequencies, thus suggesting a light-induced enhancement of inter-bilayer superconducting coupling. Correspondingly, the real part of the optical conductivity, $\sigma_1(\omega)$, develops a peak around 50 cm$^{-1}$ (Fig. 4(a)). According to our previous studies [4, 6], those features can be well fitted by assuming that 20% of the sample has been turned into a transient superconducting state with a stronger inter-bilayer coupling. Such light-enhanced low-frequency coherence arises at the expense of the spectral weight of the high-frequency transverse plasma mode [6]. A similar effect is also detected in YBa$_2$Cu$_3$O$_{6.5}$ in the normal state (Fig. 4(b)), where the measured $\sigma_2(\omega)$ enhancement has been interpreted in terms of a light-induced inter-bilayer coupling extending far above equilibrium $T_c$ [6].

In contrast with these findings, Figure 4(c) and (d) show that no low-frequency conductivity change could be detected for Bi$_2$Sr$_2$Ca$_2$Cu$_3$O$_{10}$, at all measured temperatures and time delays, although the transient spectral weight loss at the transverse plasma mode is comparable to the YBa$_2$Cu$_3$O$_{6.5}$ case.

Note that the distance between the closely spaced CuO$_2$ planes in YBa$_2$Cu$_3$O$_{6.5}$ and Bi$_2$Sr$_2$Ca$_2$Cu$_3$O$_{10}$ are similar, while the inter-bilayer separation is significantly shorter than the inter-trilayer separation (8.4 Å vs. 12 Å) [21,29]. Hence, the tunneling of Cooper pairs may take longer time in the inter-trilayer junction in Bi$_2$Sr$_2$Ca$_2$Cu$_3$O$_{10}$ than in the case of inter-bilayer tunneling in YBa$_2$Cu$_3$O$_{6.5}$. Typical time scales can be estimated from the frequencies of the plasma edge in Figures 1(b) and 1(f), which are ~30 cm$^{-1}$ for YBa$_2$Cu$_3$O$_{6.5}$ and $\lesssim$ 5 cm$^{-1}$ for $Bi_2Sr_2Ca_2Cu_3O_{10}$. These correspond to tunneling times of ~1 ps and ≳ 7 ps, respectively. Hence, the short lifetime (~3 ps) of the spectral redistribution may not be sufficient for the two pairs of trilayer units to become coherent and the respective phases to be correlated in $Bi_2Sr_2Ca_2Cu_3O_{10}$. This observation further underscores the need for long pulse mid-infrared excitation to sustain the transient state, an effort that at this stage is still frustrated by current laser technology.

However, we should also notice that the trilayer structure differs from the bilayer one in many other respects, for example, the inequivalent $CuO_2$ planes and the absence of chain oxygen. The transient structural rearrangements reported for $YBa_2Cu_3O_{6.5}$ [7], which include a decrease in the Cu-apical O distances and an increase in O-Cu-O buckling, leading to an effective hole doping to the planar Cu states, may not be reproduced in the case of $Bi_2Sr_2Ca_2Cu_3O_{10}$. In general, our results indicate that a simple stacked Josephson junction structure may not be enough for the modeling of the light-enhanced interlayer coherence in multilayer cuprates.

In summary, we compared the c-axis transient optical response of bilayer $YBa_2Cu_3O_{6.5}$ and trilayer $Bi_2Sr_2Ca_2Cu_3O_{10}$ under the same phonon pumping conditions. Both cuprates show a transient spectral weight loss of the transverse plasma mode, suggesting a weakening of the intra-bi(tri)layer coupling strength. For $YBa_2Cu_3O_{6.5}$, this results in a light-induced strengthening of the inter-bilayer coherence for $YBa_2Cu_3O_{6.5}$, while for $Bi_2Sr_2Ca_2Cu_3O_{10}$ no enhancement in inter-trilayer coherence was found. Since $YBa_2Cu_3O_{6.5}$ and $Bi_2Sr_2Ca_2Cu_3O_{10}$ share many similarities in their stacked Josephson junction structure, apical oxygen phonons, and equilibrium optical properties, their different dynamical response may suggest that additional aspects have to be taken into account for theoretical modeling of light-enhanced interlayer coherence in high-$T_c$ cuprates.

**Acknowledgements:** We gratefully acknowledge Y.-L. Mathis for support at the IR1 beam line of the synchrotron facility ANKA at the Karlsruhe Institute of Technology. We thank Jun-ichi Okamoto for discussion.

FIGURES

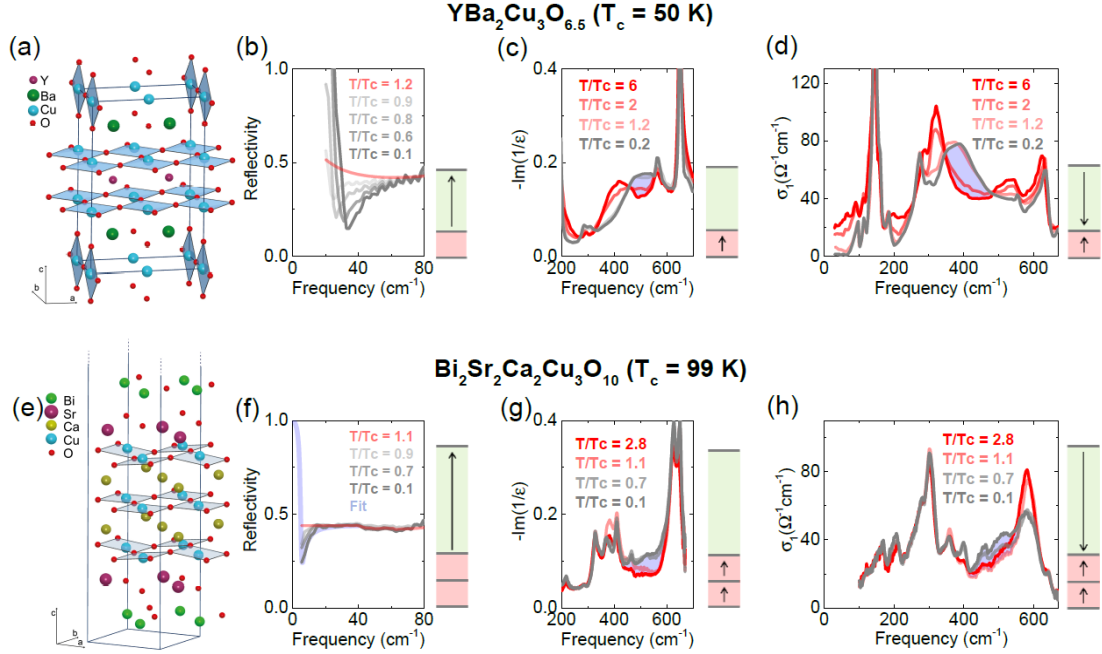

**Figure 1. (Color online) Multilayer cuprates: crystal structure and c-axis optical response at equilibrium.** Crystal structures of (a) bilayer $YBa_2Cu_3O_{6.5}$ and (e) trilayer $Bi_2Sr_2Ca_2Cu_3O_{10}$, displaying different $CuO_2$ layers and apical oxygen atoms per unit cell. Equilibrium reflectivity (b,f), energy loss function (c,g), and optical conductivity (d,h) of both samples, displayed in different frequency ranges to highlight the inter-bi(tri)layer, intra-bi(tri)layer, and transverse Josephson plasma modes, respectively. Cartoons show the currents involved for each of these modes [23]. Data in (c,d) are taken from Ref.[30], while those in (g,h) have been measured with ellipsometry on the same $Bi_2Sr_2Ca_2Cu_3O_{10}$ sample at the infrared beamline of the synchrotron radiation source ANKA in Karlsruhe. Reflectivities in (b,f) were determined by time-domain THz spectroscopy in the same setup used for our pump-probe measurements. The purple line in (f) is a fit to the data with a Josephson Plasma Resonance at $\omega = 5$ cm$^{-1}$.

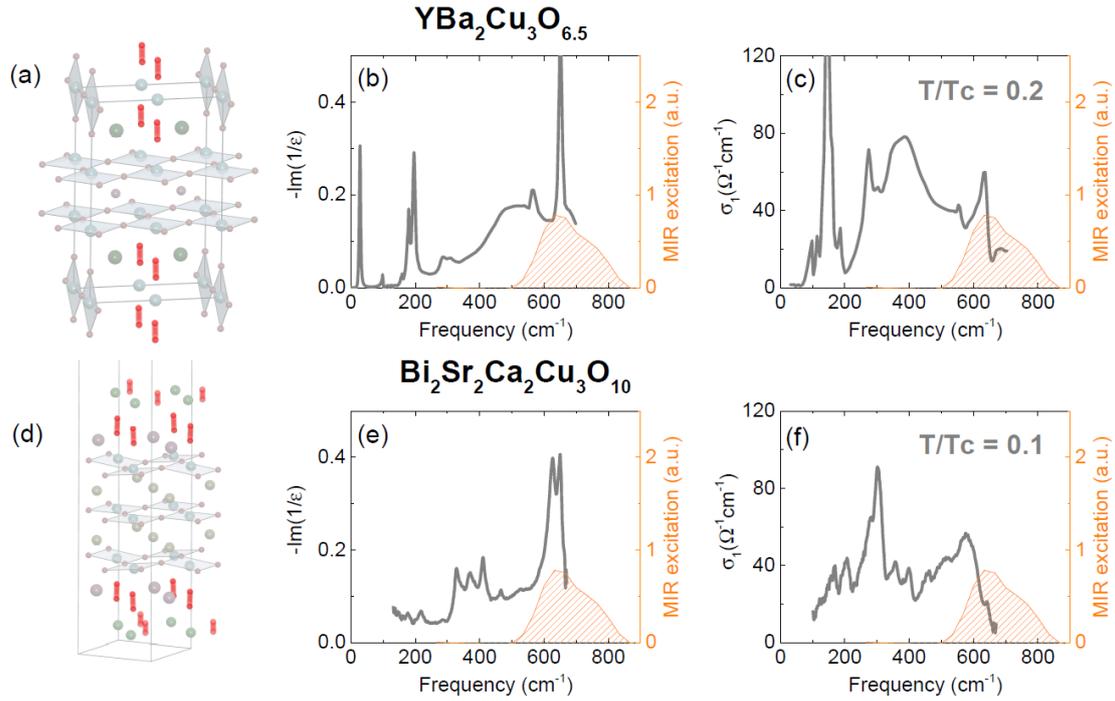

**Figure 2. (Color online) Apical oxygen excitation in YBa$_2$Cu$_3$O$_{6.5}$ and Bi$_2$Sr$_2$Ca$_2$Cu$_3$O$_{10}$.** (a,d) Cartoons of oxygen motions [31, 20] under mid-infrared resonant excitation. Energy loss function ((b) and (e)) and the real part of optical conductivity ((c) and (f)) are displayed for both cuprates at the lowest temperature, along with the spectrum of the mid-infrared pump pulses (orange).

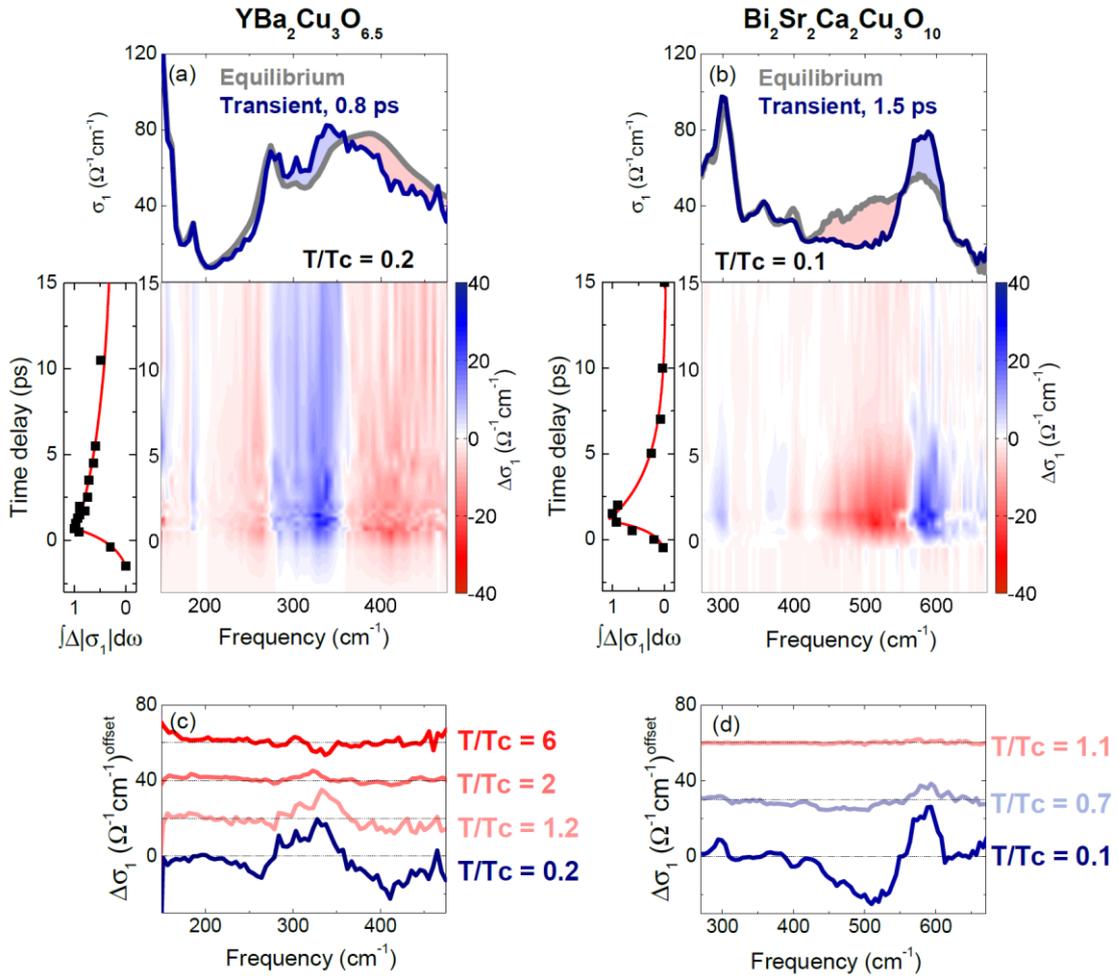

**Figure 3. (Color online) Dynamical redistribution of interlayer coherence: the high-frequency response.** The real part of the optical conductivity at equilibrium (grey) and at the maximum transient response (dark blue), measured at the lowest temperature, are shown for both samples in panels (a,b). Frequency-resolved light-induced changes (Δσ$_1$) are also displayed as color plots throughout their dynamical evolution. The transient spectral weight redistribution around the transverse plasma mode, quantified as $\int \Delta|\sigma_1(\omega)|d\omega$, shows a longer lifetime in YBa$_2$Cu$_3$O$_{6.5}$ (7 ps) than in Bi$_2$Sr$_2$Ca$_2$Cu$_3$O$_{10}$ (3 ps). Temperature-dependent measurements are displayed in panels (c,d). Data for YBa$_2$Cu$_3$O$_{6.5}$ are adapted from Ref. [6].

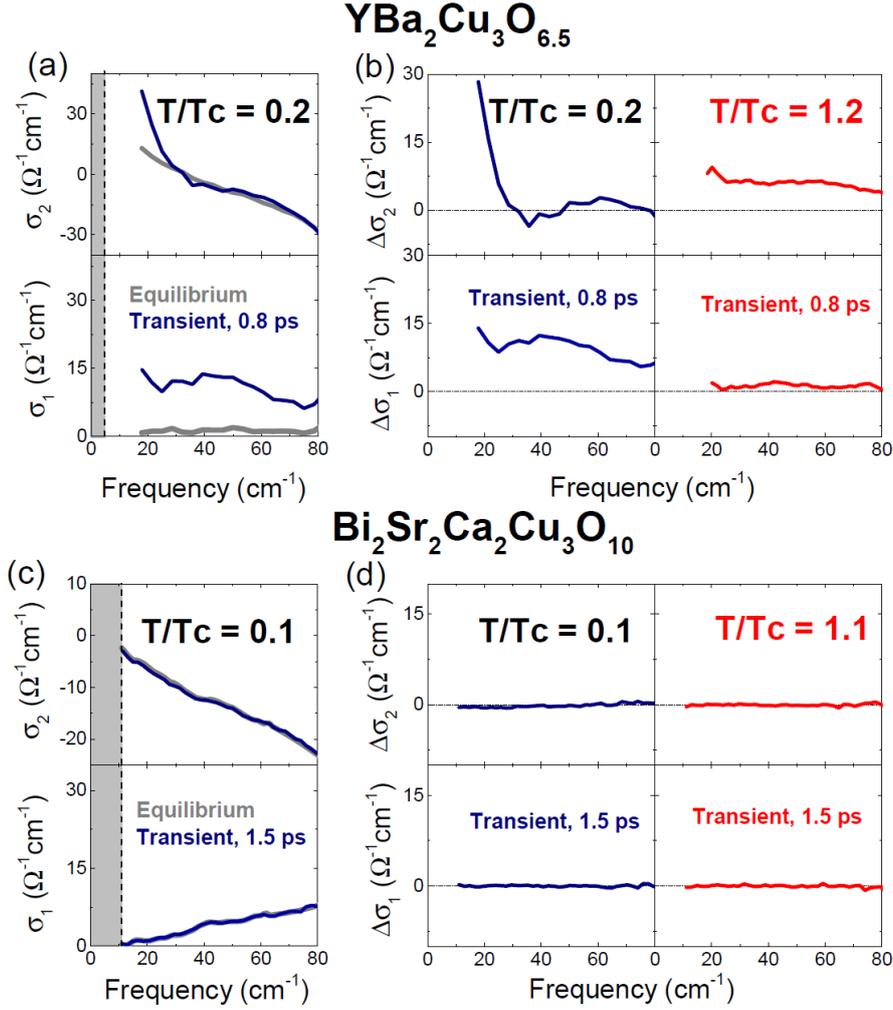

**Figure 4. (Color online) Dynamical redistribution of interlayer coherence: the low-frequency response.** Complex optical conductivities at equilibrium (grey) and at the peak of the light-induced response (dark blue) for (a) $YBa_2Cu_3O_{6.5}$ and (c) $Bi_2Sr_2Ca_2Cu_3O_{10}$, measured at $T < T_c$. The finite lifetimes of the transient states (7 ps and 3 ps, respectively) set frequency cut offs for the transient optical response (outside the shaded areas). Differential complex optical conductivity, $\Delta\sigma_{1,2} = \sigma_{1,2}^{transient} - \sigma_{1,2}^{equilibrium}$, for (b) $YBa_2Cu_3O_{6.5}$ and (d) $Bi_2Sr_2Ca_2Cu_3O_{10}$ at $T < T_c$ (dark blue) and $T > T_c$ (red).

[1] D. Fausti, R. I. Tobey, N. Dean, S. Kaiser, A. Dienst, M. C. Hoffmann, S. Pyon, T. Takayama, H. Takagi, A. Cavalleri, Science 331, 189 (2011).

[2] C. R. Hunt, D. Nicoletti, S. Kaiser, T. Takayama, H. Takagi, and A. Cavalleri, Phys. Rev. B 91, 020505(R) (2015).

[3] M. Först, R. I. Tobey, H. Bromberger, S. B. Wilkins, V. Khanna, A. D. Caviglia, Y.-D. Chuang, W. S. Lee, W. F. Schlotter, J. J. Turner, M. P. Minitti, O. Krupin, Z. J. Xu, J. S. Wen, G. D. Gu, S. S. Dhesi, A. Cavalleri, and J. P. Hill, Phys. Rev. Lett., 112, 157002 (2014)

[4] S. Kaiser, C. R. Hunt, D. Nicoletti, W. Hu, I. Gierz, H. Y. Liu, M. Le Tacon, T. Loew, D. Haug, B. Keimer, and A. Cavalleri, Phys. Rev. B 89, 184516 (2014).

[5] M. Först, A. Frano, S. Kaiser, R. Mankowsky, C. R. Hunt, J. J. Turner, G. L. Dakovski, M. P. Minitti, J. Robinson, T. Loew, M. Le Tacon, B. Keimer, J. P. Hill, A. Cavalleri, and S. S. Dhesi

Phys. Rev. B, 90, 184514 (2014).

[6] W. Hu, S. Kaiser, D. Nicoletti, C. R. Hunt, I. Gierz, M. C. Hoffmann, M. Le Tacon, T. Loew, B. Keimer, and A. Cavalleri, Nature Mater. 13, 705 (2014).

[7] R. Mankowsky, A. Subedi, M. Först, S.O. Mariager, M. Chollet, H. Lemke, J. Robinson, J. Glownia, M. Minitti, A. Frano, M. Fechner, N. A. Spaldin, T. Loew, B. Keimer, A. Georges, A. Cavalleri, Nature 516, 71 (2014).

[8] M. Mitrano, A. Cantaluppi, D. Nicoletti, S. Kaiser, A. Perucchi, S. Lupi, P. Di Pietro, D. Pontiroli, M. Riccò, S. R. Clark, D. Jaksch, and A. Cavalleri, Nature 530, 461 (2016).

[9] Zachary M. Raines, Valentin Stanev, and Victor M. Galitski, Phys. Rev. B 91, 184506 (2015).

[10] Aavishkar A. Patel and Andreas Eberlein, Phys. Rev. B 93, 195139 (2016).

[11] R. Höppner, B. Zhu, T. Rexin, A. Cavalleri, and L. Mathey, Phys. Rev. B 91, 104507 (2015).

[12] Jun-ichi Okamoto, Andrea Cavalleri, Ludwig Mathey, Phys. Rev. Lett. 117, 227001 (2016).

[13] S. J. Denny, S. R. Clark, Y. Laplace, A. Cavalleri, and D. Jaksch, Phys. Rev. Lett. 114, 137001 (2015).

[14] M. A. Sentef, A. F. Kemper, A. Georges, and C. Kollath, Phys. Rev. B 93, 144506 (2016).

[15] Michael Knap, Mehrtash Babadi, Gil Refael, Ivar Martin, Eugene Demler, Phys. Rev. B 94, 214504 (2016).

[16] M. Kennes, E. Y. Wilner, D. R. Reichman, and A. J. Millis, Nature Physics (2017) DOI: 10.1038/NPHYS4024

[17] A. Trokiner, L. Le Noc, J. Schneck, A. M. Pougnet, R. Mellet, J. Primot, H. Savary, Y. M. Gao, and S. Aubry, Phys. Rev. B 44, 2426(R) (1991); B. W. Statt and L. M. Song, Phys. Rev. B 48, 3536 (1993).

[18] S. Ideta, K. Takashima, M. Hashimoto, T. Yoshida, A. Fujimori, H. Anzai, T. Fujita, Y. Nakashima, A. Ino, M. Arita, H. Namatame, M. Taniguchi, K. Ono, M. Kubota, D. H. Lu, Z.-X. Shen, K. M. Kojima, and S. Uchida, Phys. Rev. Lett. 104, 227001 (2010).

[19] E. J. Singley, M. Abo-Bakr, D. N. Basov, J. Feikes, P. Guptasarma, K. Holldack, H. W. Hübers, P. Kuske, Michael C. Martin, W. B. Peatman, U. Schade, and G. Wüstefeld, Phys. Rev. B 69, 092512 (2004).

[20] N. N. Kovaleva, A. V. Boris, T. Holden, C. Ulrich, B. Liang, C. T. Lin, B. Keimer, C. Bernhard, J. L. Tallon, D. Munzar, and A. M. Stoneham, Phys. Rev. B 69, 054511 (2004).

[21] A. V. Boris, D. Munzar, N. N. Kovaleva, B. Liang, C. T. Lin, A. Dubroka, A. V. Pimenov, T. Holden, B. Keimer, Y.-L. Mathis, and C. Bernhard, Phys. Rev. Lett. 89, 277001 (2002).

[22] Y. Hirata, K. M. Kojima, M. Ishikado, S. Uchida, A. Iyo, H. Eisaki, and S. Tajima, Phys. Rev. B 85, 054501 (2012).

[23] D. van der Marel, A. A. Tsvetkov, Phys. Rev. B 64, 024530 (2001); D. van der Marel, A. Tsvetkov, Czech. J. Phys. 46, 3165 (1996).

[24] A. Dubroka, M. Rössle, K. W. Kim, V. K. Malik, D. Munzar, D. N. Basov, A. A. Schafgans, S. J. Moon, C. T. Lin, D. Haug, V. Hinkov, B. Keimer, Th. Wolf, J. G. Storey, J. L. Tallon, and C. Bernhard, Phys. Rev. Lett. 106, 047006 (2011).

[25] Adam Dubroka, Dominik Munzar, Physica C 405, 133 (2004).

[26] I.-C. Ho, X. Guo, and X.-C. Zhang, Opt. Express 18, 2872 (2010).

[27] R. Huber et al., App. Phys. Lett. 76, 3191 (2000).

[28] Martin Dressel and George Grüner, Electrodynamics of Solids,

[29] D. Munzar, C. Bernhard, A. Golnik, J. Humlíček, M. Cardona, Solid State Commun. 112, 365 (1999).

[30] C. C. Homes, T. Timusk, D. A. Bonn, R. Liang, W. N. Hardy, Physica C 254, 265 (1995).

[31] C. C. Homes, T. Timusk, D. A. Bonn, R. Liang, W. N. Hardy, Can. J. Phys. 73, 663 (1995).